\documentclass{PoS}

\usepackage{graphicx}
\usepackage{amsmath,amssymb,mathrsfs}
\usepackage{fancyhdr}
\usepackage{psfrag}
\usepackage{array,bbm}
\usepackage{url}
\usepackage{rotating}
\usepackage{multirow}
\usepackage{dsfont}
\usepackage{cancel}

\newenvironment{Eqnarray}%
     {\arraycolsep 0.14em\begin{eqnarray}}{\end{eqnarray}}
\newcommand{\beqa}{\begin{Eqnarray}}
\newcommand{\eeqa}{\end{Eqnarray}}
\newcommand{\beq}{\begin{equation}}
\newcommand{\eeq}{\end{equation}}

\title{The CP-conserving 2HDM after the 8 TeV run}

\ShortTitle{2HDM after the 8 TeV run}

\author{P. M. Ferreira\\
        Instituto Superior de Engenharia de Lisboa- ISEL, 1959-007 Lisboa, Portugal and \\
        Centro de F\'{\i}sica Te\'{o}rica e Computacional, Faculdade de Ci\^{e}ncias, Universidade de Lisboa,
        Av.\ Prof.\ Gama Pinto 2, 1649-003 Lisboa, Portugal \\
        E-mail: \email{ferreira@cii.fc.ul.pt}}

\author{Renato Guedes\\
        Centro de F\'{\i}sica Te\'{o}rica e Computacional, Faculdade de Ci\^{e}ncias, Universidade de Lisboa,
        Av.\ Prof.\ Gama Pinto 2, 1649-003 Lisboa, Portugal \\
        E-mail: \email{renato@cii.fc.ul.pt}}

\author{John F. Gunion\\
        Davis Institute for High Energy Physics,
    University of California,
    Davis, California 95616, USA \\
        E-mail: \email{gunion@physics.ucdavis.edu}}

\author{Howard E.~Haber\\
 Santa Cruz Institute for Particle Physics,
    University of California and \\
       Ernest Orlando Lawrence Berkeley National Laboratory,
University of California, Berkeley, California 94720, USA \\
        E-mail: \email{haber@scipp.ucsc.edu}}

\author{Marco O. P. Sampaio\\
        Departamento deF\'{\i}sica da Universidade de Aveiro and I3N
        Campus de Santiago, 3810-183 Aveiro, Portugal\\
        E-mail: \email{msampaio@ua.pt}}

\author{\speaker{Rui Santos}\\
        Instituto Superior de Engenharia de Lisboa- ISEL, 1959-007 Lisboa, Portugal and \\
        Centro de F\'{\i}sica Te\'{o}rica e Computacional, Faculdade de Ci\^{e}ncias, Universidade de Lisboa,
        Av.\ Prof.\ Gama Pinto 2, 1649-003 Lisboa, Portugal \\
        E-mail: \email{rsantos@cii.fc.ul.pt}}

\abstract{
We confront the most common CP-conserving 2HDM with the LHC data analysed so far
while taking into account all previously available experimental data. A special allowed corner of the parameter
space is analysed - the so-called wrong-sign scenario where the Higgs coupling to down-type quarks changes 
sign relative to the Standard Model while the coupling to the massive vector bosons does not.
}

\FullConference{XXII. International Workshop on Deep-Inelastic Scattering and Related Subjects,\\
		28 April - 2 May 2014\\
		Warsaw, Poland}

\begin{document}

\section{Introduction}
The end of the 8 TeV run at the Large Hadron Collider (LHC) has confirmed the existence of a Higgs boson~\cite{ATLASHiggs, CMSHiggs} that
very much resembles the one predicted by the Standard Model (SM). Furthermore there are no hints of
extra scalars in the data analysed so far. As many of the extension of the SM, the two-Higgs doublet model (2HDM)
is being cornered into a SM-like region except for a few regions of the parameter space.

The 2HDM is an extension of the SM where one extra doublet is added to the particle content
while keeping the SM symmetries. It appears in the literature in a variety of versions that depend mainly on the extra symmetries
imposed on the Lagrangian and on how the $SU(2) \times U(1)$ symmetry is broken to $U(1)$. In this work we will focus
on the CP-conserving 2HDM with a $Z_2$ discrete symmetry, softly broken in the potential by a dimension two term.

We will show the allowed parameter space of the model after the LHC 8 TeV run with all theoretical and experimental
constraints taken into account. We will then study a region of the parameter space where the lightest CP-even Higgs (considered to be the
SM-like Higgs throughout the paper)  coupling to the down-type
quarks changes sign relative to the SM.

\section{The allowed parameter space of the 2HDM}

%%%%%%%%%%%%%%%%%%%%%%%%%%%%%% minimal version of the model section

The most general 2HDMs give rise to couplings corresponding to tree-level Higgs-mediated flavour-changing neutral 
currents (FCNCs), in clear disagreement with experimental data. A simple and natural way to 
avoid tree-level FCNCs is to impose a $Z_2$ symmetry on the
scalar doublets, $\Phi_1 \rightarrow \Phi_1$,
$\Phi_2 \rightarrow - \Phi_2$,  and a corresponding symmetry on the
quark fields. This leads to the well known four Yukawa model types I, II, Flipped (F) (or Y)
and Lepton Specific (LS) (or X). The different Yukawa types are built such that 
only $\phi_2$ couples to all fermions (type I), or $\phi_2$ couples to up-type quarks and $\phi_1$ couples to 
down-type quarks and leptons (type II), or $\phi_2$ couples to up-type quarks and 
to leptons and $\phi_1$ couples to down-type quarks (type F) or finally $\phi_2$ couples to all 
quarks and $\phi_1$ couples to leptons (type LS). See~\cite{hhg} for a comprehensive review on the 2HDM.

The scalar potential in a softly broken $Z_2$ symmetric 2HDM can be written as
\begin{align*}
V(\Phi_1,\Phi_2) =& m^2_1 \Phi^{\dagger}_1\Phi_1+m^2_2
\Phi^{\dagger}_2\Phi_2 - (m^2_{12} \Phi^{\dagger}_1\Phi_2+{\mathrm{h.c.}
}) +\frac{1}{2} \lambda_1 (\Phi^{\dagger}_1\Phi_1)^2 +\frac{1}{2}
\lambda_2 (\Phi^{\dagger}_2\Phi_2)^2\nonumber \\ 
+& \lambda_3
(\Phi^{\dagger}_1\Phi_1)(\Phi^{\dagger}_2\Phi_2) + \lambda_4
(\Phi^{\dagger}_1\Phi_2)(\Phi^{\dagger}_2\Phi_1) + \frac{1}{2}
\lambda_5[(\Phi^{\dagger}_1\Phi_2)^2+{\mathrm{h.c.}}] ~, \label{higgspot}
\end{align*}
where $\Phi_i$, $i=1,2$ are complex SU(2) doublets. We will focus on a specific realisation
of the 2HDM, the usual 8-parameter CP-conserving potential where the potential parameters and the VEVs are all real.
In this model we choose as free parameters, the four masses, the rotation
angle in the CP-even sector, $\alpha$, the ratio of the vacuum expectation
values,  $\tan\beta=v_2/v_1$, and the soft breaking parameter $m_{12}^2$.
By convention, we take $0\leq\beta\leq \pi/2$ and $- \pi/2 \leq  \alpha \leq \pi/2$.

The 2HDM parameters are chosen such that electric charge is conserved
while neutral Higgs fields acquire real vacuum expectation values.
Note that the existence of a tree-level scalar potential minimum that breaks the electroweak symmetry but 
preserves both the electric charge and CP symmetries, ensures that no additional tree-level potential minimum
that spontaneously breaks the electric charge and/or CP symmetry
can exist~\cite{vacstab}. Further, we force the CP-conserving minimum to be the global one~\cite{Barroso:2013awa}.

In order to find the 2HDM parameter space that is still allowed after the 8 TeV run we have used ScannerS~\cite{Coimbra:2013qq}
interfaced with SusHi~\cite{Harlander:2012pb} and HDECAY~\cite{Djouadi:1997yw, Harlander:2013qxa}
for Higgs production and decays, cross-checked with HIGLU~\cite{Spira:1995mt} and 2HDMC~\cite{Eriksson:2009ws}.
The remaining Higgs production cross sections were taken from~\cite{LHCHiggs}. All collider data was taken into
account with HiggsBounds~\cite{Bechtle:2013wla} and HiggsSignals~\cite{Bechtle:2013xfa}. The
remaining constraints (see~\cite{Barroso:2013zxa}), theoretical, electroweak precision and B-physics constraints are coded in ScannerS.

We have performed a scan in the 2HDM parameter space in the following range: $m_h = 125.9~{\rm GeV}$, 
$m_h + 5~{\rm GeV} <m_H,\, m_A < 1~{\rm TeV}$, $100~{\rm GeV} < m_{H^\pm} < 1~{\rm TeV}$, 
$1 < \tan \beta < 30$, $|\alpha| < \pi/2$
and  $- (50~{\rm GeV})^2<m_{12}^2 <(500~{\rm GeV})^2$. 
\begin{figure}[h!]
\centering
\includegraphics[width=5.0in,angle=0]{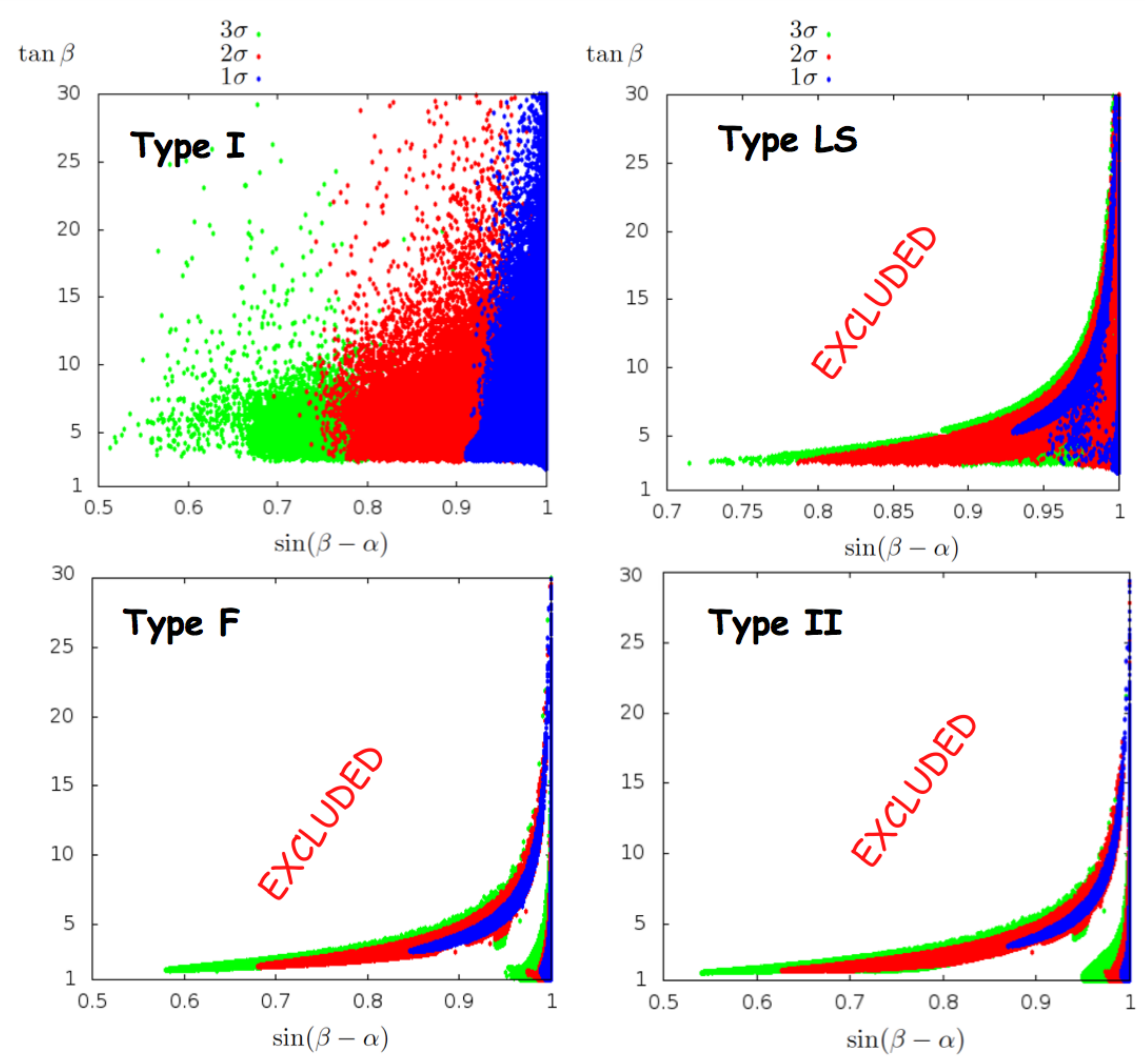}
\caption{Allowed parameter space for models I, LS, F and II after the 8 TeV run.}
\label{fig:fig1}
\end{figure}
\vspace{-0.1cm}
In figure~\ref{fig:fig1} we present the allowed
parameter space after the 8 TeV run at $1\sigma$, $2\sigma$ and $3\sigma$ with all experimental and theoretical
constraints taken into account. There are some interesting features worth mentioning. 
First, the bounds on $\sin(\beta - \alpha)$ range from about $0.5$ in type II to about $0.7$ in type LS at  $3\sigma$.
Second, that large values of $\tan \beta$ require $\sin (\beta -\alpha)$ close to 1 except for the type I model. 
Note that although the Higgs couplings to quarks are equal in types I and LS, the couplings to leptons are different. 
As a result, the measurement of $pp \to h \to \tau^ + \tau^ -$ affects considerably more
the parameter space of type LS than that of type I~\cite{Arhrib:2011wc}.
Finally,
it is clear from the figure that in models type II and Flipped, the allowed points are centred around two lines. 
The line on the right corresponds to the SM-like limit, that is $\sin (\beta - \alpha) =1$. In this limit, the lightest Higgs 
couplings to gauge bosons and to fermions are the SM ones. The line on the left corresponds to the limit  
$\sin(\beta +\alpha)=1$. In type II and with our conventions, it corresponds to the limit where the Higgs coupling to down-type
quarks changes sign relative to the SM while couplings to up-type quarks and massive gauge bosons are the same.
This is called the wrong-sign limit~\cite{Ferreira:2014naa} (see also~\cite{Dumont:2014wha, Fontes:2014tga} . 
Note that this limit is imposed only at tree-level.

\section{The wrong-sign scenario}

We will now analyse the wrong-sign scenario in the light of the next run of the LHC. This
scenario was first discussed in~\cite{Ginzburg:2001ss}. Let us start by defining $\kappa_i^2=\Gamma^{\scriptscriptstyle {\rm 2HDM}}  (h \to i)/\Gamma^{\scriptscriptstyle {\rm SM}}  (h \to i)$
which means that at tree-level $\kappa_i$ is just the ratio of couplings $\kappa_i=g^{\scriptscriptstyle {\rm 2HDM}}  /g^{\scriptscriptstyle {\rm SM}} $. In the SM-like limit,
$\kappa_{W(Z)} = \sin (\beta - \alpha) =1$, implies $\kappa_U=\kappa_D=\kappa_L =1$, that is, the lightest Higgs couplings
to up-type quarks ($U$),  down-type quarks ($D$) and leptons ($L$), are the SM ones. The wrong-sign scenario can be defined
as either $\kappa_D \, \kappa_W <0$ or $\kappa_U \, \kappa_W <0$ ($\kappa_L$ never plays a major role in the interference terms).
We can further have $\kappa_D \, \kappa_U <0$, in which case both $\kappa_g$ and $\kappa_\gamma$ can be affected or 
$\kappa_D \, \kappa_U >0$ meaning that only $\kappa_\gamma$ can be affected.
The wrong sign scenario is obtained in type II and F with
\begin{equation}
\sin (\beta + \alpha) = 1 \Rightarrow \kappa_D = -1 \, (\kappa_U = 1); \quad \sin(\beta - \alpha) =\frac{\tan^2 \beta -1}{\tan^2 \beta +1} \Rightarrow \kappa_W>0 \, (\tan \beta > 1) .
\label{eq:eq1}
\end{equation}
%while for all four types it can also be obtained in the limit 
%\begin{equation}
%\sin (\beta + \alpha) = -1 \Rightarrow \kappa_U = -1; \quad \sin(\beta - \alpha) =-\frac{\tan^2 \beta -1}{\tan^2 \beta +1} \Rightarrow \kappa_W>0 \, (\tan \beta < 1) \, .
%\label{eq:eq2}
%\end{equation}
As the constraints from B-physics and from the $R_b$ measurements force $\tan \beta$ to be above $O(1)$ we will focus on the case $\kappa_D=-1$.
We will now discuss this scenario for the type II model (very similar to the type F case).

Will the LHC be able to probe the wrong-sign scenario? We define the signal strength as
\begin{equation}
\mu^h_f \, = \, \frac{\sigma \, {\rm BR} (h \to
  f)}{\sigma^{\scriptscriptstyle {\rm SM}} \, {\rm BR^{\scriptscriptstyle{\rm SM}}} (h \to f)}
\label{eg-rg}
\end{equation}
where $\sigma$ is the Higgs production cross section and ${\rm BR} (h \to f)$ is
the branching ratio of the decay into some given final state $f$;  $\sigma^{\scriptscriptstyle {\rm {SM}}}$
and ${\rm BR^{\scriptscriptstyle {\rm SM}}}(h \to f)$ are the expected values for the same quantities in the SM. 
We will not separate different LHC initial state production mechanisms and instead sum over all production mechanisms in computing the cross section.  

As a rough approximation of the precision with current data we require that the LHC's $\mu^h_f$ for the final states 
$f=WW$, $ZZ$, $b\bar b$, $\gamma \gamma$ and $\tau^+ \tau^ -$  
are each consistent with unity within 20\%. In order to understand how an increase in precision 
will affect the wrong-sign scenario we then require that all the $\mu^h_f$ are within 10\% or 5\% of the SM prediction.
We now have to answer the following questions. First, why isn't this scenario excluded yet? Second, will it be probed
at the LHC with high energy and high luminosity?

In the limit $\sin (\beta + \alpha) \to 1$, $\kappa_D \to -1$ which implies that the main Higgs production mode, 
gluon-gluon fusion, is enhanced. The quark initiated modes are not modified relative to the SM because there are no 
interference terms (at LO). VBF and associated production do not suffer any significant change because 
$\sin (\beta - \alpha) \approx \sin (\beta + \alpha)$ for $\tan \beta \gg 1$. Therefore the scenario could in
principle be probed at the production level in $gg \to h$ due to the interference between top and bottom loops.
However, the uncertainties in the gluon fusion process advise not to use this production process
to distinguish between the two scenarios~\cite{Ferreira:2014naa,Fontes:2014tga}. 

Regarding the Higgs decays, it is clear that there is no difference between the two scenarios in the decay to fermions. Again,
because $\sin (\beta - \alpha) \approx \sin (\beta + \alpha)$ for $\tan \beta \gg 1$, taking $\tan \beta = 8$ the ratio
of the wrong sign $\Gamma (h \to WW \, (ZZ))$ decay width to the respective SM width is 0.94, which corresponds
to a negligible effect in $\mu^h_{WW} $ due to the already discussed Higgs production cross section enhancement.
Therefore, we have to turn into
the decays where the interference between the different loop contributions occur, that is $h \to \gamma \gamma$
and $h \to gg$.  
\begin{figure}[h!]
\centering
\includegraphics[width=3.02in,angle=0]{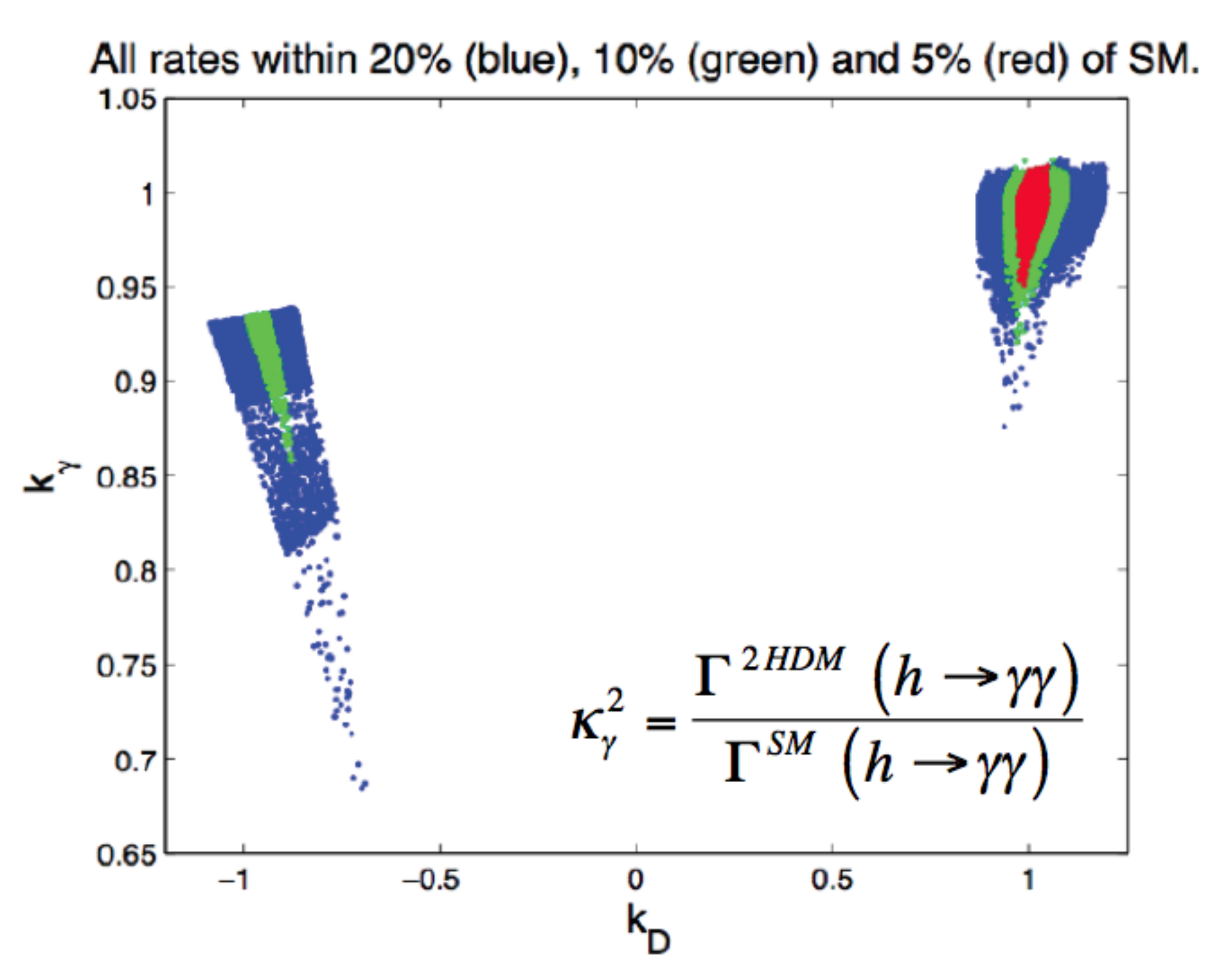}
\hspace{-.3cm}
\includegraphics[width=2.97in,angle=0]{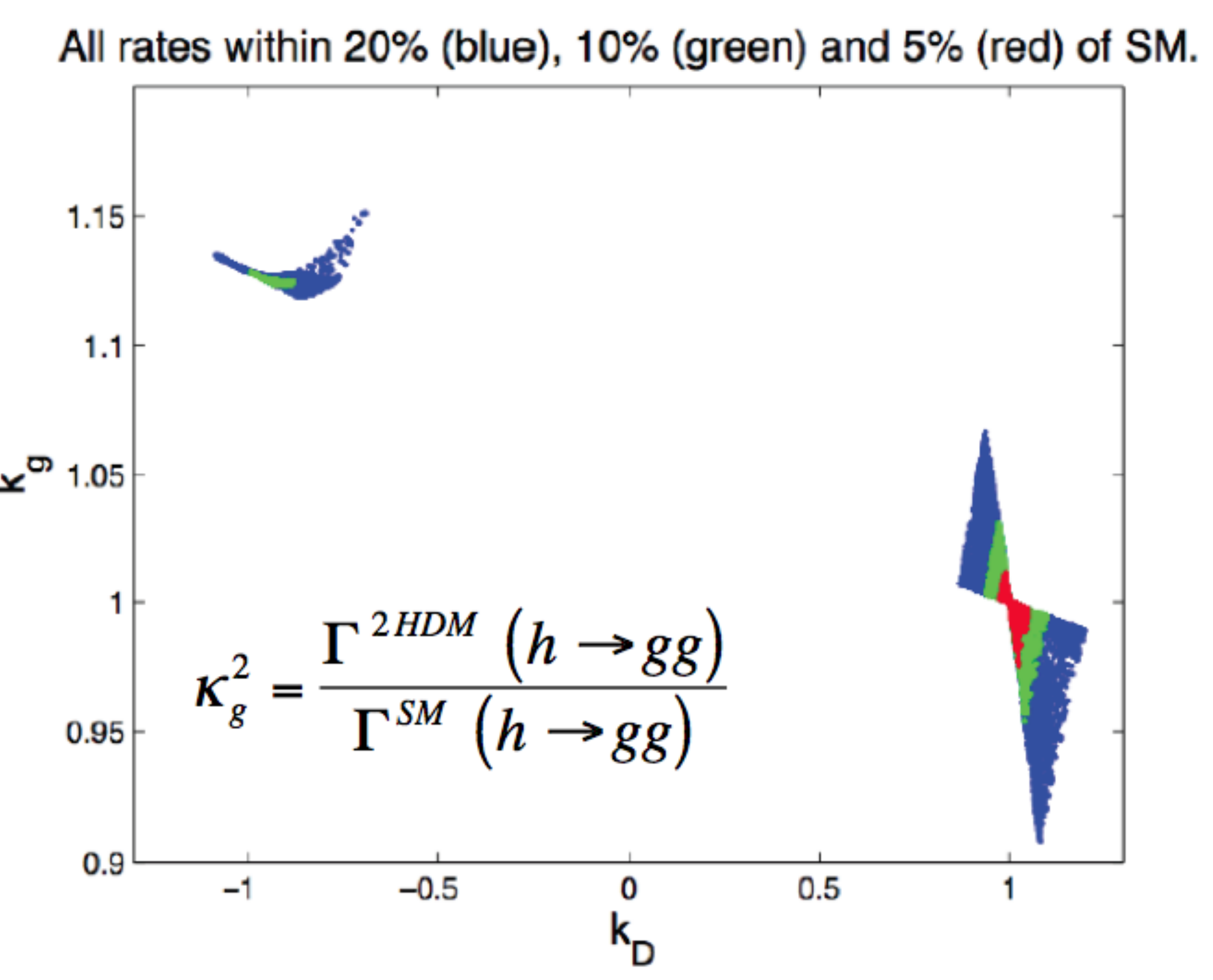}
\caption{Left panel: $\kappa_\gamma$ as a function of $\kappa_D$; right panel: $\kappa_g$ as a function of $\kappa_D$ }
\label{fig:T4}
\end{figure}
In figure \ref{fig:T4} we present $\kappa_\gamma$ (left) and $\kappa_g$ (right) as a function of $\kappa_D$ in type II with all rates within
$20\%$ (blue), $10\%$ (green) and $5\%$ (red) of the SM values. While it is understandable that a measurement of $\kappa_g$
could probe the wrong-sign scenario, the same is not true for $\kappa_\gamma$. In fact, the decay $h \to \gamma \gamma$
has not only the top and bottom loop contributions but also the W and charged-Higgs ones. Neglecting the charged-Higgs contribution,
 a change in the sign of $\kappa_D$ amounts to about a 1\% difference in the width. Therefore it is the charged Higgs loop that is responsible for the more 
substantial reduction in $\Gamma (h \to \gamma \gamma)$. This effect is due to the nondecoupling of the
charged Higgs loop. As shown in~\cite{Ferreira:2014naa} the charged-Higgs contribution to $\Gamma (h \to \gamma \gamma)$
in the $\kappa_D <0$ case is approximately constant and always sufficiently significant as to eventually be observable at the LHC. However,
the constraints coming from tree-level unitarity imply that the result is only perturbatively reliable for $m_{H^\pm} \sim 650$ GeV.

According to Table~1-20 of Ref.~\cite{Dawson:2013bba}, the expected errors for $\kappa_g$ based on fittings are 
$6$--$8\%$ for $L=300$ fb$^{-1}$ and $3$--$5\%$ for $L=3000$ fb$^{-1}$ (for 14 TeV). The predicted
accuracy for $\kappa_\gamma$ is $5$--$7\%$ for an integrated luminosity of $L=300$ fb$^{-1}$ and $2$--$5\%$ for $L=3000$ fb$^{-1}$.
Therefore there are good chances to probe the wrong-sign scenario in the 14 TeV LHC run. Also with the predicted accuracy
for the International Linear Collider~\cite{Ono:2012ah, Asner:2013psa} the scenario could not only be probed by 
a measurement of $\kappa_g$  and $\kappa_\gamma$ but also in the process $e^+ e^- \to Z h ( \to b \bar{b}) $. Finally one should 
mention that a thorough study of this scenario has to take into account the 2HDM electroweak corrections, some
of which are already available~\cite{LopezVal:2009qy, Kanemura:2014dja}. \\[0.05cm]

\large{\textbf{Acknowledgments}}
PMF, RG and RS are supported by FCT
under contracts PTDC/FIS
/117951/2010 and PEst-OE/FIS/UI0618/2011.
RG is also supported by a FCT Grant SFRH/BPD/47348/2008.
MS is supported by a FCT Grant SFRH/BPD/69971/2010.
%The work of J.F.G. and H.E.H. is supported
%in part by the U.S. Department of Energy, under grant
%numbers DE-SC-000999 and DE-FG02-04ER41268, respectively.

%\section{Conclusions}

\end{document}